# IMPROVED PROBABILISTIC SEISMIC DEMAND-INTENSITY RELATIONSHIP: HETEROSCEDASTIC APPROACHES

## Libo Chen

College of Civil Engineering, Fuzhou University, Fuzhou, Fujian, P.R. China

**Abstract**

As an integral part of assessing the seismic performance of structures, the probabilistic seismic demand-intensity relationship has been widely studied. In this study, the phenomenon of heteroscedasticity in probabilistic seismic demand models was systematically investigated. A brief review of the definition, diagnosis, and conventional treatment of heteroscedasticity is presented herein, and based on that, two more generalized methods for both univariate and multivariate cases are proposed. For a typical four-span simply supported girder bridge, a series of nonlinear time history analyses were performed through multiple stripe analysis to determine its seismic demand-intensity that can be employed as a sample set. For both univariate and multivariate cases, probabilistic seismic demand models were developed based on the two aforementioned methods under the Bayesian regression framework, and the fitted results were compared and analyzed with the conventional models using linear regression approaches. In the presence of probabilistic seismic demand considering heteroscedasticity, the patterns of non-constant variance or covariance can be characterized effectively, and a better-calibrated prediction region than that of homoscedastic models can be provided. The causes of the heteroscedasticity phenomenon and subsequent solutions are thoroughly discussed. The analysis procedures can be further embedded in seismic fragility and risk assessment, thus providing a more accurate basis for aseismic decision-making.

**Keywords**

Heteroscedasticity; probabilistic seismic demand; multiple-stripe analysis; intensity measure; Bayesian; covariance regression

## 1. INTRODUCTION

In the past two decades, the performance-based earthquake engineering (PBEE) framework proposed by the Pacific Earthquake Engineering Research Center has played a significant role in seismic design and risk assessment for different engineering structures. One of the critical links to be investigated in this field is the probabilistic seismic demand analysis, which is mainly used to establish a probabilistic relationship between the structural seismic demand and ground motion intensity. Typically, a series of nonlinear time-history analyses is performed for a specific structure. Alternative analysis procedures are available in the literature for characterizing the relationship between engineering demand parameter (EDP) and intensity measure (IM) based on recorded ground motions, such as the cloud analysis method, multiple-stripe analysis (MSA), and incremental dynamic analysis (IDA). The seismic demand values of the critical components are recorded and then matched with the input ground motion intensities to generate the EDP-IM dataset. Probabilistic seismic demand modeling can be summarized as a regression problem using statistical approaches. Several researchers have conducted extensive research on this issue and subsequently applied it to seismic fragility analysis and risk assessment.

Following the earlier studies, it is usually assumed that the probabilistic relationship between seismic demand, $D$ and spectral acceleration, $S_a$ can be expressed as the power model[1,2]:

$$D = \alpha_0 (S_a)^{\alpha_1} e \tag{1}$$

where $\alpha_0$ and $\alpha_1$ are the regression parameters, $e$ is the random error term that can be represented by a lognormal distribution with a unit median. The logarithmic transformation of Equation 1 is used as the demand model.

$$\ln(D) = \ln(\alpha_0) + \alpha_1 \ln(S_a) + \sigma\varepsilon \tag{2}$$

Under the assumption of homoscedasticity ($\sigma^2$ is constant and independent of $S_a$) and normality ($\varepsilon$ follows the standard normal distribution), a simple linear regression can be performed to estimate the equation parameters including $\alpha_0$, $\alpha_1$, and $\sigma$.

Although this classic seismic demand model has been extensively used in subsequent studies, few limitations have been indicated by researchers. For instance, Jalayer reported that the dispersion measure of the seismic demand tends to increase for larger values of spectral acceleration[3]; thus, assuming the variance as a constant in the linear model might be non-conservative at some instances. Baker compared the estimates of log standard deviation using the simple cloud and stripe methods for all spectral acceleration levels, and similar conclusions were obtained.[4] It also stresses the importance of performing linear regressions locally for some cases, i.e., over a narrower range of spectral acceleration values. Freddi et al. evaluated the homoscedasticity assumption for all demand models, and observed that this condition was not applicable to local EDPs, regardless of IM.[5] Thus, they concluded that the variability of the dispersion should be considered when defining the fragility curves of RC building components.

To address the aforementioned problems, piecewise (segmented) linear regression approaches were introduced. Mackie and Stojadinović first attempted the bilinear least-squares fitting to generate probabilistic seismic demand models for typical highway overpass bridges.[6] Ramamoorthy et al. and Simon et al. found that the linear model cannot accurately represent the relationship between seismic demand and intensity measure in log-space[7,8]; it tends to underestimate the demand for small and large values of $S_a$ and overestimates the demand for intermediate values of $S_a$ for specific targets. In response to this observation, a bilinear regression model was developed that can provide a better fit over the entire range of $S_a$. The bilinear model can be expressed as follows:

$$\ln(D) = \begin{cases} \theta_{01} + \theta_{11} \ln(S_a) + \sigma_1 \varepsilon_1 & \ln(S_a) \leq \theta_{S_a} \\ (\theta_{01} + \theta_{11}\theta_{S_a}) + \theta_{21}[\ln(S_a) - \theta_{S_a}] + \sigma_2 \varepsilon_2 & \ln(S_a) > \theta_{S_a} \end{cases} \tag{3}$$

where $\theta_{01}$ and $\theta_{11}$ are the ordinate intercept and slope of the first branch of the model, respectively; $\theta_{21}$ is the slope of the second branch; $\theta_{S_a}$ is the inflection point along the $\ln(S_a)$ axis; $\sigma_1$ and $\sigma_2$ are the constant standard deviations of the model error for each branch, respectively; $\varepsilon_1$ and $\varepsilon_2$ are random variables that follow a standard normal distribution. Recently, O'Reilly derived the bilinear demand-intensity relationship minutely under the existing SAC/FEMA closed-form framework.[9]

Nevertheless, from the author's point of view, the bilinear model possesses some disadvantages, summarized as follows: First, the inflection point is not apparent in many situations, and it is typically designated by the subjective judgment, but not derived from data, which can introduce some additional error into the inference of the model; second, the homoscedasticity assumption is still held at each branch and the variances estimated from the two branches of data are significantly different, which will lead to a discontinuity in the subsequent fragility analysis at the inflection point; and third, the phenomenon of heteroscedasticity at each branch can still be observed in many cases, specifically for the second branch, which also indicates that more segments should be used through piecewise approach.

Further, Aslani and Miranda used a function to represent variation in the logarithmic standard deviation of structural response parameters with changes in ground motion intensity,[10] which can be expressed as follows:

$$\tilde{\sigma}_r = \beta_1 + \beta_2(IM) + \beta_3(IM)^2 \tag{4}$$

where $IM$ is the ground motion intensity measure; $\tilde{\sigma}_r$ is the logarithmic standard deviation of the demand, evaluated at a given $IM$; $\beta_1$, $\beta_2$, and $\beta_3$ are the parameters computed from a regression analysis with the known $IM - \tilde{\sigma}_r$ pairs. Huang et al. observed the non-constant standardized residuals by the diagnostic plots for the shear models after using natural logarithmic transformation[11]; thus, a weighted least-squares regression was used to handle the remaining heteroscedasticity of the models. The former obtained the patterns by regressing the samples of statistical characteristics from the data sets once again, while the latter did not elaborate this issue. Existing research lacks in developing effective means to represent and solve the problem for practical probabilistic seismic demand assessment.

Additionally, the joint distribution of multiple seismic demand parameters is a concern in some cases. When performing a probabilistic seismic evaluation of structures, multiple demand parameter values for the same component could be evaluated. For instance, it may be essential to know the joint distribution of displacement ductility and residual displacement for a pier column of a bridge under near-fault ground motions. Another possible scenario is that when developing a seismic fragility model for a system, we need to investigate the seismic demand values of different components simultaneously, and then build a joint probabilistic seismic demand model for subsequent analysis. A common approach is to make an additional assumption that the multiple seismic demand values follow a multivariate lognormal distribution; then, a multivariate linear regression (MLR) analysis[12] can be used to investigate the relationship between the multiple seismic demand parameters and ground motion intensity measure in the log space.

$$\ln(\boldsymbol{D}) = \boldsymbol{\beta_0} + \boldsymbol{\beta_1} \ln(IM) + \boldsymbol{\epsilon} \tag{5}$$

Where, $\ln(\boldsymbol{D}) = \left(\ln(D_1), \ln(D_2) \ldots \ln(D_p)\right)^{\mathrm{T}}$ is a vector of seismic demand responses, $\boldsymbol{\beta_0} = \left(\beta_{01}, \beta_{02} \ldots \beta_{0p}\right)^{\mathrm{T}}$ and $\boldsymbol{\beta_1} = \left(\beta_{11}, \beta_{12} \ldots \beta_{1p}\right)^{\mathrm{T}}$ are two vectors of unknown parameters to be estimated, and the error term $\boldsymbol{\epsilon} = \left(\epsilon_1, \epsilon_2 \ldots \epsilon_p\right)^{\mathrm{T}}$ has $\mathrm{E}(\boldsymbol{\epsilon}) = \boldsymbol{0}$ and $\mathrm{Cov}(\boldsymbol{\epsilon}) = \boldsymbol{\Sigma}$. Similarly, the least-squares method can be used to make related statistical inferences. In addition to estimating the mean of $\ln(\boldsymbol{D})$, MLR provides information on the residual variance and correlation between the different demand parameters. By adopting this method, some researchers have established joint probabilistic seismic demand models for different types of structures.[13,14] However, the homoscedasticity assumption still holds for this model, i.e., the covariance matrix of the residuals is assumed to be a constant matrix. This assumption is probably inconsistent with the actual situation. A natural conjecture is whether the correlations of the seismic demand residuals are IM dependent or not. To the best of the author's knowledge, the research on investigating the multivariate heteroscedasticity in probabilistic seismic demand modeling has never been conducted and reported previously.

This study mainly focused on the modeling of the variance and covariance in probabilistic seismic demand analysis. The models for both the mean and variance or covariance were specified, and the relevant parameters were estimated simultaneously using Bayesian approaches. The portrayal of the variance was performed while making statistical inferences about the mean, thus providing a better description of the problem. Section 2 briefly reviews the basic theory of heteroscedasticity, including its definition, testing methods, and conventional treatments. Two more generalized solution approaches are presented for the univariate and multivariate problems. In Section 3, a typical reinforced concrete girder bridge is used as an example to establish its nonlinear finite element model, and the selection and scaling schemes for ground motion records are discussed, based on which a series of nonlinear dynamic time analyses is implemented to develop the required IM-EDP dataset for demand analysis. In Section 4, the development of univariate and multivariate probabilistic seismic demand models by employing the two aforementioned methods considering heteroscedasticity is discussed, and the results are compared with those of classical linear regression models. Finally, summary of this study and conclusions are presented in Section 5.

**2. METHODOLOGY**

*2.1 Diagnostic and conventional treatment*

In classical regression, the constant variance assumption states that, given the predictor variables, the conditional variance of the error term is consistent for all observations. In other words, whenever the predictor variables vary, the corresponding response exhibits the same variance around the regression line. This is called homoscedasticity. In contrast, the condition of the error variance not being constant in overall observations is called heteroscedasticity.

An informal method for detecting heteroscedasticity is a graphical method in which regression analysis is performed assuming that there is no heteroscedasticity, and then a visual inspection of residuals plotted against fitted values or any of the explanatory variables is conducted to check if these exhibit any systematic pattern. Furthermore, several formal methods can be used to test for the presence of heteroscedasticity in a regression model. With some prior knowledge about the source of heteroscedasticity, the Breusch–Pagan test can be used to test whether the error variances from the regression are dependent on the values of the independent variables, derived from the Lagrange multiplier test principle.[15] Another popular test method is the White test, which is a general approach for testing heteroscedasticity in the error distribution by regressing the squared residuals on all distinct covariates, the cross-products of covariates, and the squares of covariates.[16] These two test methods have been widely used in econometrics over the past few decades.[17]

If heteroscedasticity exists, two general approaches can be used to address the problem. First is the use of transformation, in which the response is transformed to homoscedasticity. The most commonly used family is the modified power transformations of Box and Cox,[18] in which powers and logarithm of responses are parameterized by a number $\lambda$. The $\lambda$ can be estimated by maximizing the likelihood. However, it should be noted that this family of transformations leads to tractable math, which does not eliminate the need to run all the diagnostic checks after the transformation. In fact, some researchers have indicated that a single transformation cannot necessarily produce the normality, constancy of variance, and linearity of systematic effects for the mean and dispersion models simultaneously.[19] Consider a set of Poisson variables $y_i \sim Possion(\mu_i), i = 1,2,\ldots,n$; to equalize variances, the transformation $z = y^{1/2}$ is needed, to produce (approximate) normality $z = y^{2/3}$, and to produce additivity $z = \ln(y)$. This is a special case, but it also indicates that this issue still needs to be addressed in some cases if heteroscedasticity exists for the data after the transformation.

Weighted least squares (WLS) is another approach that can be used when the least-squares assumption of constant variance in the residuals is violated. The use of the WLS approach will solve the problem of bias in the standard errors and also provide more efficient estimates. With the correct weight, this procedure minimizes the sum of the weighted squared residuals to produce residuals with constant variance (homoscedasticity). However, determining a proper weight to be used can be a challenging task. Generally, observations with small variances should have relatively large weights, and vice versa. The ideal weight is the reciprocal of the variance of the error, but this is usually incalculable, and other approaches, such as feasible generalized least squares,[17] may be needed.

*2.2 Univariate heteroscedasticity*

A standard way to consider the variance that depends systematically on the level of response or some other factor is to postulate a formal model for response variance,[20] as one models the mean response. Possible heteroscedasticity can be incorporated by specifying a variance function, $g$. The function $g$ depends on the mean response, addition's parameter vector $\gamma$, and some other explanatory variables $z_i$, which may include some or all of the components of $x_i$. A very general specification for the mean and variance is

$$E(y_i) = \mu_i = f(x_i, \beta), \ Var(y_i) = \sigma^2 g^2(\mu_i, z_i, \gamma). \tag{6}$$

Over the decades, several researchers have conducted a series of specific studies on this issue. In particular, Harvey proposed a model of multiplicative heteroscedasticity,[21] which was a very flexible model comprising most of the useful formulations as special cases. In this study, Harvey's model was selected to construct a probabilistic seismic demand model for a single component. The general formulation is as follows:

$$\begin{cases} y_i \sim \mathcal{N}(\mu_i, \sigma_i^2), \\ \mu_i = x_i^T \beta, \quad i = 1,2,\ldots,n \\ \ln \sigma_i^2 = z_i^T \gamma \end{cases} \tag{7}$$

where $y_i, i = 1,2,\ldots,n$ is the observed response on the $i$-th value; $x_i = (x_{i1}, \ldots, x_{ip})^T$ and $z_i = (z_{i1}, \ldots, z_{iq})^T$ are the explanatory variables $X = (X_1, \ldots, X_p)^T$ and $Z = (Z_1, \ldots, Z_q)^T$, respectively; $\beta = (\beta_1, \ldots, \beta_p)^T$ is a $p \times 1$ vector of unknown parameters for the mean model, and $\gamma = (\gamma_1, \ldots, \gamma_q)^T$ is a $q \times 1$ vector of unknown parameters for the variance model. $z_i$ may contain some or all of the variables in $x_i$ and the variables that are not included in $x_i$, and $z_i$ is assumed to contain a constant 1. The log-linear form ensures that $\sigma_i^2$ remains positive.

Harvey discussed maximum likelihood estimation and the likelihood ratio test for the model.[21] Later, Aitkin provided an alternate iterative algorithm for the model to obtain maximum likelihood estimates for the parameters.[22] Subsequently, Cepeda and Gamerman implemented a Bayesian approach to estimate the model parameters.[23]

*2.3 Multivariate heteroscedasticity*

Historically, models for multivariate heteroscedasticity have been developed in the context of multivariate time series, for which various multivariate "autoregressive conditionally heteroscedastic" models have been studied in the econometric literature.[24] However, the applicability of such models was limited to situations in which heteroscedasticity was temporal. Chiu et al. suggested to model the logarithm of the covariance as a linear function of the explanatory variables[25]; however, the interpretation of the model parameters was significantly complicated. Pourahmadi presented a Cholesky decomposition of the covariance matrix, which allowed for a meaningful interpretation of the parameters.[26] However, it was suitable only when a natural ranking of responses existed, similar to that in longitudinal or time series studies. Hoff and Niu proposed a covariance regression model that directly models the covariance matrix as a function of the explanatory variables.[27] It was a natural extension of the mean regression model, and the parameters could be interpreted in a way similar to those in a mean regression. This model was utilized to construct the joint probabilistic seismic demand model in this study. The detailed description is as follows:

Let $y$ be a $p \times 1$ vector of multivariate responses and $x$ be a $q \times 1$ vector of explanatory variables. The mean and covariance matrices of $y$ are represented as $\mu_x = E[y \mid x]$ and $\Sigma_x = \text{Cov}[y \mid x]$, respectively. The simplest version of the covariance regression model expresses $\Sigma_x$ as $\Sigma_x = \Psi + Bxx^T B^T$, where $\Psi$ is a $p \times p$ positive definite matrix, and $B$ is a $p \times q$ real matrix that has a direct interpretation in terms of how heteroscedasticity co-occurs among the $p$ variables of $y$. Additionally, the model also has an interpretation as a type

of random-effects model. Assume that the observed data $y_1, \ldots, y_n$ has the following model:

$$\begin{aligned}
y_i &= \mu_{x_i} + \gamma_i \times Bx_i + \epsilon_i \\
E[\epsilon_i] &= 0, \text{Cov}[\epsilon_i] = \Psi \\
E[\gamma_i] &= 0, \text{Var}[\gamma_i] = 1, E[\gamma_i \times \epsilon_i] = 0
\end{aligned} \qquad (8)$$

The resulting covariance matrix for $y_i$ given $x_i$ is then

$$\begin{aligned}
E\left[(y_i - \mu_{x_i})(y_i - \mu_{x_i})^T\right] &= E[\gamma_i^2 Bx_i x_i^T B^T + \gamma_i(Bx_i \epsilon_i^T + \epsilon_i x_i^T B^T) + \epsilon_i \epsilon_i^T] \\
&= Bx_i x_i^T B^T + \Psi \\
&= \Sigma_{x_i}
\end{aligned} \qquad (9)$$

The model specified by Equation 8 restricts the difference between $\Sigma_x$ and the baseline matrix $\Psi$ to be a rank-one matrix. Further flexibility can be gained by adding additional random effects, allowing the difference between $\Sigma_x$ and the baseline matrix $\Psi$ to be of any desired rank up to and including $p$. The random-effects representation for a rank-$r$ covariance regression model is

$$\begin{aligned}
y_i &= \mu_{x_i} + \sum_{k=1}^{r} \gamma_{i,k} \times B^{(k)} x_i + \epsilon_i \\
&= \mu_{x_i} + \tilde{B}(\gamma_i \otimes x_i) + \varepsilon_i, \text{ where } \tilde{B} = \left(B^{(1)}, \ldots, B^{(r)}\right)
\end{aligned} \qquad (10)$$

Two methods of statistical inference for this model were provided in the literature.[27] One of these is maximum likelihood estimation via the expectation–maximization (EM) algorithm which is straightforward, and another is Bayesian estimation using a Gibbs sampler to approximate the posterior distribution.

## 3. CASE STUDY

*3.1 Project background and finite element model*

A simply-supported reinforced concrete box girder bridge, designed as per Chinese standards, as shown in Figure 1, was selected as a comprehensive case study.[28] It consists of four spans, and the length of each span 20 m. The superstructure is constructed by C40 concrete. The widths of the top and bottom plates of the box girder are 8.5 and 4.5 m, respectively, and the height of the girder is 1.3 m. The substructure of the bridge adopts a single-column pier with a circular cross-section. The height and diameter of the piers are 9.0 and 1.3 m, respectively. Eighteen HRB335 steel bars with a diameter of 25 mm are used as the longitudinal reinforcement in the piers. HPB235 steel bars with a diameter of 12 mm are used as stirrups with a spacing of 0.1 m. The thickness of the concrete cover is 0.04 m. The cap beams are 6.0 m long, with a rectangular cross-section of 2.5 × 1.2 m². C30 concrete is used for both the piers and cap beams. Pile-supported abutments and a bored pile foundation are provided. Eight plate-rubber bearings with dimensions of 300 mm × 500 mm × 84 mm (length, width, and thickness) are placed between the main girder and cap beams, set in two rows.

A three-dimensional finite element model was established in the OpenSees platform to simulate the nonlinear dynamic response of the bridge.[29] Existing research findings suggest that for such a bridge, the beam components mostly remain elastic when subjected to seismic events, while the bearings and the substructure are damaged. Herein, elastic and force-based beam-column elements were adopted to model the beam and substructure, respectively. Further, Concrete02 material was selected to simulate the constitutive relationship of the concrete in the substructure. With the varying peak compressive stress and softening slope, the model could consider the contribution of transverse stirrups. The hysteretic material was selected to simulate the behavior of the steel reinforcement used in the substructure. This could reflect the pinching effect of the reinforcement stress and strain, as well as the damage caused by ductility and energy. For the abutment, the hyperbolic gap material was selected to simulate the constraining effect of passive pressure of the backfill after pounding occurs between the girder and abutment on the longitudinal displacement of the abutment. In addition, the hysteretic material was selected to describe the resistance of the pile foundation to the longitudinal displacement of the abutment. A transverse rigid bar of the same length as the width of the deck was used. The nonlinear springs connected in series to the gap element were used to model the passive backfill response and expansion joint. The plate-rubber bearings were simulated using the elastomeric bearing element, which considers the shear deformation and slip behavior of the bearing. To improve the computational efficiency, a linear spring element was used to simulate the pile-soil effect of the bridge. The general layout and description of the bridge model are presented in Figure 1.

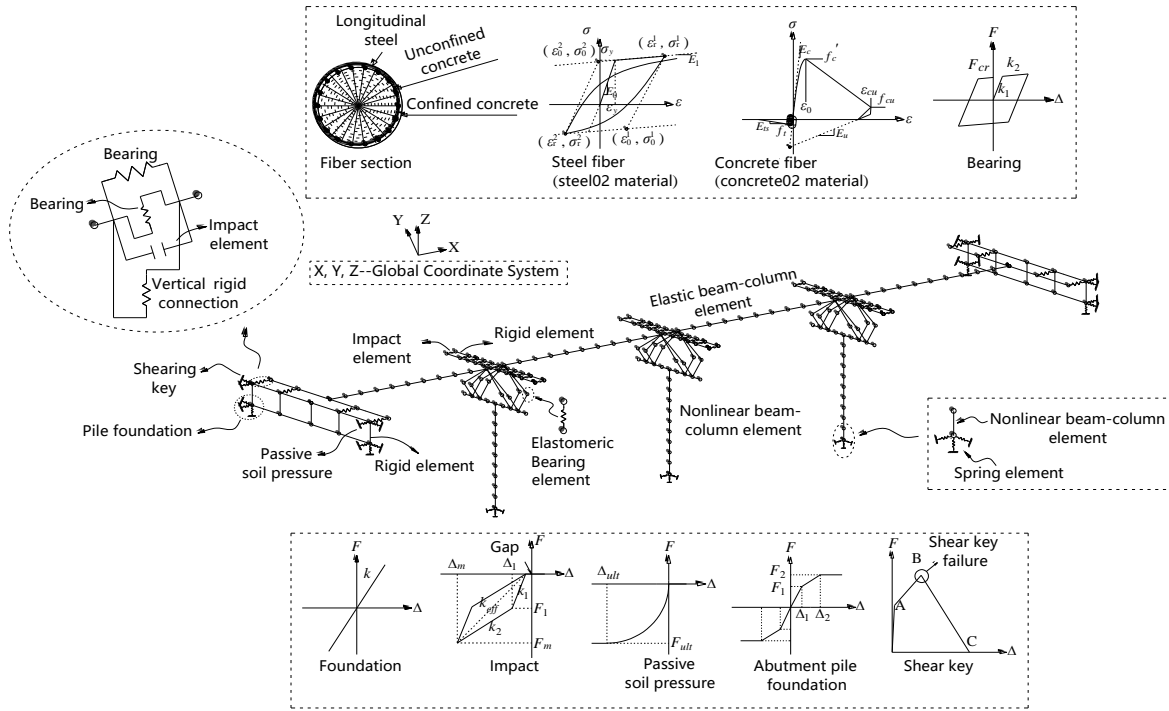

Figure 1 General layout and description of the bridge model

*3.2 Selection and scaling of ground motion records*

The seismic response of the bridge system is affected by uncertainties in the earthquake input (record-to-record variability), properties defining the system (model parameter uncertainty), and lack of knowledge (epistemic uncertainty).[30] Previous studies have shown that generally the ground-motion uncertainty is significantly larger than the numerical seismic response modeling uncertainty. Accordingly, in this study, only the effect of record-to-record variability was considered in the numerical simulation. Based on the full consideration of the uncertainty of ground motion, 80 seismic records were selected from the Pacific earthquake engineering research (PEER) database.[31] The principles of seismic record selection are as follows:

(1) The moment magnitudes of earthquake events range between 6.0–8.0;
(2) The epicenter distance ranges between 15–35 km;
(3) The shear wave velocity of the soil layer of 30 m thickness below the surface satisfies $260 m/s \leq V_{S30} \leq 510 m/s$, which is consistent with the soil type II of Chinese Specification (the soil type for the site of the case study bridge);

(4) Aftershock and pulse-like records were not considered in this study.

In addition, it is necessary to scale the selected ground motion records to achieve the desired levels of seismic intensity. Amplitude scaling of the seismic record to the target earthquake intensity may cause a change in the shape of the earthquake spectrum, resulting in a difference between the structural responses generated by the amplitude-scaling ground motion record and the real ground motion record at that seismic intensity, which affects the evaluation of the seismic performance of the structure. Therefore, it is necessary to strictly control the magnitude of the amplitude scaling factor to avoid excessive changes in the shape of response spectrum of the records. Recently, several researchers have studied the deviation caused by the amplitude scaling of ground motions and recommended the value of the amplitude scaling coefficient.[32,33] For the ground motion records selected in this study, the corresponding amplitude scaling coefficient was strictly controlled between 0.25-4.0.

*3.3 Multiple-stripe analysis*

In this study, MSA was adopted for the nonlinear time history analysis of the bridge.[32] Its basic idea was to amplify or reduce the selected series of ground motion records to a specific intensity level by amplitude scaling, and then record the seismic response of the structural component under the seismic intensity to form a "strip" of the seismic demands at that intensity level. For the range of ground motion intensities of interest, this operation can be repeated to amplify the ground motion records to different intensity levels to obtain "strips" at multiple intensity levels. Compared with the cloud analysis method, MSA can more conveniently estimate the mean and standard deviation of logarithmic seismic demand samples under a certain seismic intensity, and the probabilistic characteristics of seismic demand under this intensity level can be obtained according to these two statistical parameters.

Previous studies have shown that seismic demand parameters satisfy the lognormal distribution assumption.[5,10] This assumption was retained in the subsequent analysis and the regression analysis was conducted in logarithmic space. In previous studies, most researchers used an equidistant amplitude scaling scheme of ground motion intensity in natural coordinates for selected ground motions when performing MSA. To ensure the uniformity of the distribution of the covariate (IM) in logarithmic space, an equidistant amplitude scaling processing technique was applied to 80 selected seismic records in logarithmic coordinates. The specific scheme was as follows: the logarithmic ground motion intensity $\ln Sa(T_1)$ was scaled from -2.3 to 0.1, and the step size was selected as 0.1, that is, the amplitude scaling of 80 ground motion records was conducted 25 times at equal intervals in logarithmic space, which was converted to natural coordinates from 0.100g to 1.105g. The specific amplitude-scaling scheme for the selected ground motion records is presented in Table 1.

Table 1. Amplitude scaling scheme for the selected ground motion records

| Sequence Number | $Sa$ in logarithmic coordinates /ln(g) | $Sa$ in natural coordinates /g | Sequence Number | $Sa$ in logarithmic coordinates /ln(g) | $Sa$ in natural coordinates /g | Sequence Number | $Sa$ in logarithmic coordinates /ln(g) | $Sa$ in natural coordinates /g |
|---|---|---|---|---|---|---|---|---|
| 1 | -2.3 | 0.100 | 11 | -1.3 | 0.273 | 21 | -0.3 | 0.741 |
| 2 | -2.2 | 0.111 | 12 | -1.2 | 0.301 | 22 | -0.2 | 0.819 |
| 3 | -2.1 | 0.123 | 13 | -1.1 | 0.333 | 23 | -0.1 | 0.905 |
| 4 | -2.0 | 0.135 | 14 | -1.0 | 0.368 | 24 | 0.0 | 1.000 |
| 5 | -1.9 | 0.150 | 15 | -0.9 | 0.407 | 25 | 0.1 | 1.105 |
| 6 | -1.8 | 0.165 | 16 | -0.8 | 0.449 | | | |

| | | | | | |
|---|---|---|---|---|---|
| 7 | -1.7 | 0.183 | 17 | -0.7 | 0.497 |
| 8 | -1.6 | 0.202 | 18 | -0.6 | 0.549 |
| 9 | -1.5 | 0.223 | 19 | -0.5 | 0.607 |
| 10 | -1.4 | 0.247 | 20 | -0.4 | 0.670 |

Past earthquake disasters have shown that piers, as the most critical part of girder bridges, are most likely to be damaged under earthquakes and often exhibit a strong nonlinear response characteristic. In previous studies, researchers have employed several different indicators to quantify the damage in piers under seismic action, such as drift ratios, displacement ductility, and curvature ductility. In this study, the displacement ductility of three piers was selected as the engineering demand parameter. The displacement ductility was calculated by the ratio of the recorded maximum displacement $\delta_{max}$ and yield displacement $\delta_{yield}$ (obtained by Pushover analysis, $\delta_{yield} = 0.027m$ for this case) at the top of the pier. Because this study mainly focused on the phenomenon of heteroscedasticity in the probabilistic seismic demand analysis, the seismic demands of the other components (bearings, abutments, etc.) were not recorded in the analysis. The considered IM for this case study was the elastic spectral acceleration at fundamental period (T1 = 0.9 s), because this IM has been approved as an optimal IM for probabilistic seismic demand modeling of typical short-to-medium span highway bridges.

Based on the aforementioned analysis, a total of 80 × 25 = 2000 nonlinear time history analyses were conducted for the selected bridge model. The displacement ductility ratios of the three piers and ground motion intensities were recorded and paired one by one to form the EDP-IM data set for subsequent analysis. Figure 2 shows the overall distribution of the displacement ductility ratios of the three piers under seismic action in log-log space. The black scattered points in the figure represent the original demand sample points. It can be clearly observed that the dispersion of the displacement ductility ratios increases with an increase in the ground shaking intensity, particularly for piers 1 and 3. In general, the distribution of the sample points of the three piers was "trumpet-shaped," showing different degrees of heteroscedasticity.

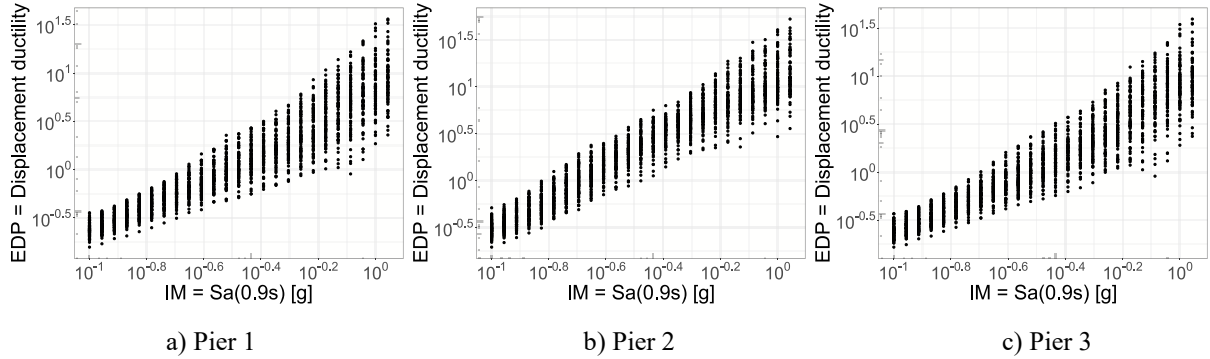

a) Pier 1     b) Pier 2     c) Pier 3

Figure 2 EDP-IM data generated through MSA

## 4. RESULTS AND COMPARISON

*4.1 Univariate heteroscedasticity*

Based on the data obtained in the previous section, the seismic demand models for each pier were established in logarithmic space. As mentioned earlier, Harvey's model was used to build a univariate heteroscedasticity model, as shown in Equation 7. To fully consider the nonlinear trend of the mean and variance under the logarithmic space, a third-order polynomial was used as the basis function for regression. Suppose $x_i = \ln IM_i, y_i = \ln EDP_i$, and $\boldsymbol{x}_i = (1, x_i, x_i^2, x_i^3)^T$. Assuming that the mean model is $E(y_i|x_i) = \boldsymbol{x}_i^T\boldsymbol{\beta}$, the variance model is $\text{Var}(y_i|x_i) = \exp(\boldsymbol{x}_i^T\boldsymbol{\gamma})$,

where $\boldsymbol{\beta} = (\beta_0, \beta_1, \beta_2, \beta_3)^T$ and $\boldsymbol{\gamma} = (\gamma_0, \gamma_1, \gamma_2, \gamma_3)^T$ are parameters to be estimated. The Markov chain Monte Carlo (MCMC) method was used to estimate the unknown parameters in a Bayesian framework, which is implemented in the probabilistic programming language Stan using R.[34] The weakly informative priors were used for the parameters, i.e., the Normal distribution $\mathcal{N}(0,100)$. Four chains were initialized with different starting values, and 5000 iterations with a thinning factor of 10 (i.e., using every 10[th] iteration) were conducted. The first half of each chain was discarded as a burn-in part. Convergence of the MCMC samples was assessed using the Brooks–Gelman–Rubin diagnostic ($\hat{R} < 1.05$) and Monte Carlo standard error (MCSE $< 0.05$).

Meanwhile, the first-order linear model was selected as the comparison model. The mean and variance models were $E(y_i|x_i) = \alpha_0 + \alpha_1 x_i$ and $\text{Var}(y_i|x_i) = \sigma^2$, respectively. The parameters $(\alpha_0, \alpha_1, \sigma)$ were estimated using the classical least-squares method.

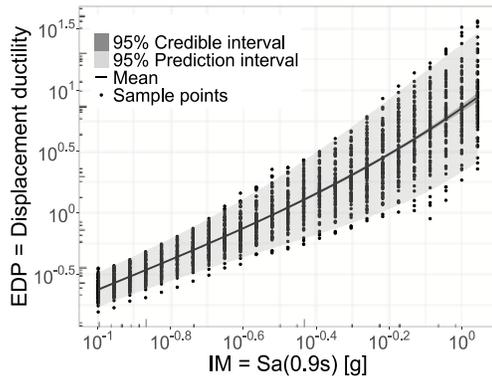

a) PSDM based on heteroscedastic model for pier 1

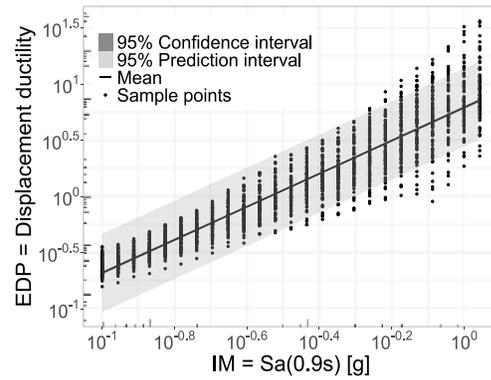

b) PSDM based on linear model for pier 1

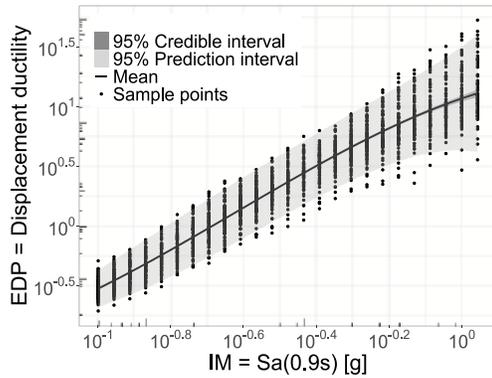

c) PSDM based on heteroscedastic model for pier 2

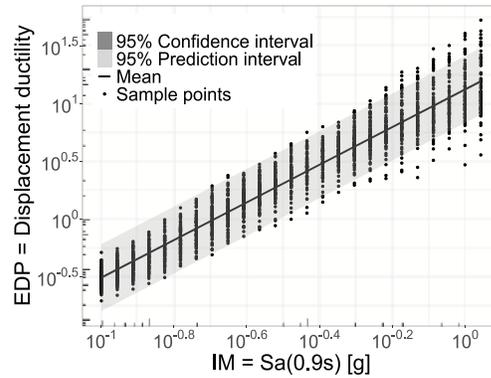

d) PSDM based on linear model for pier 2

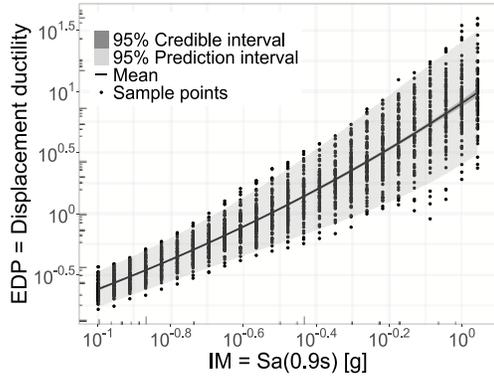
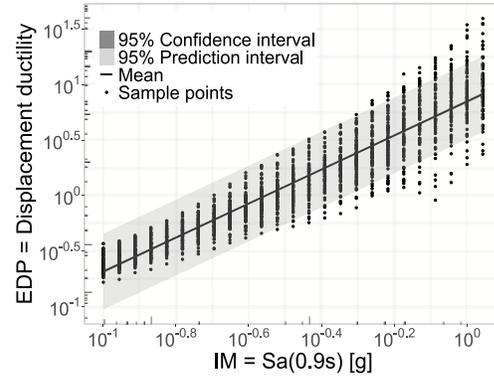

e) PSDM based on heteroscedastic model for pier 3    f) PSDM based on linear model for pier 3

Figure 3 Comparison of the overall fitting of PSDM based on the two approaches for piers 1, 2, and 3

Figure 3 shows a comparison of the probabilistic seismic demand models for the three piers based on the two approaches, where (a), (c), and (e) show the results based on the heteroscedastic model, and (b), (d), and (f) show the results based on classical linear regression. Through comparison, it was observed that the heteroscedastic model can reasonably well fit the distribution of sample points. The prediction intervals of the seismic response of the three piers tended to expand nonlinearly with the gradual increase in ground motion intensity, and the overall appearance was "trumpet-shaped." The prediction intervals of the classical linear regression model were always of the same size due to the assumption of homoscedasticity, and the overall appearance was "straight-shaped." Thus, the interval estimation results were larger when the ground motion intensity was small and vice versa. The credible intervals (confidence intervals) of both models fluctuate insignificantly, and the upper and lower bounds were close to the mean line, indicating that the mean fit of the models was good. It was evident that the probabilistic seismic demand model based on heteroscedasticity can better represent the response pattern of seismic demand and fit data with higher accuracy.

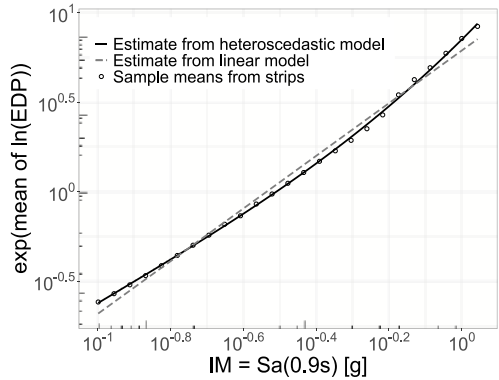
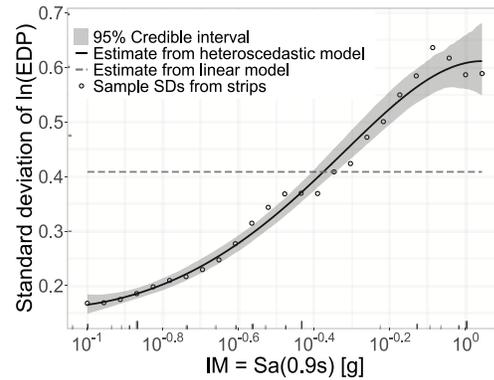

a) Mean values of $EDP|IM$ for pier 1    b) Standard deviation of $lnEDP|IM$ for pier 1

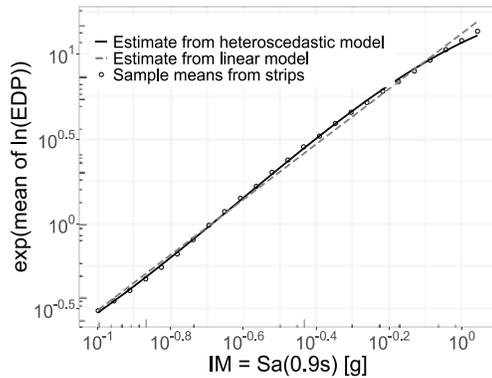
c) Mean values of $EDP|IM$ for pier 2

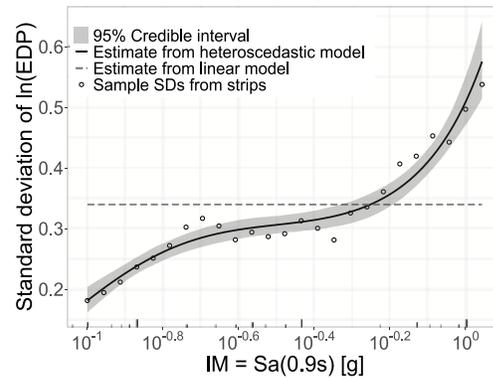
d) Standard deviation of $lnEDP|IM$ for pier 2

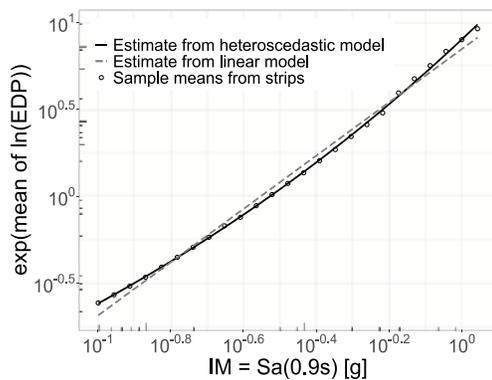
e) Mean values of $EDP|IM$ for pier 3

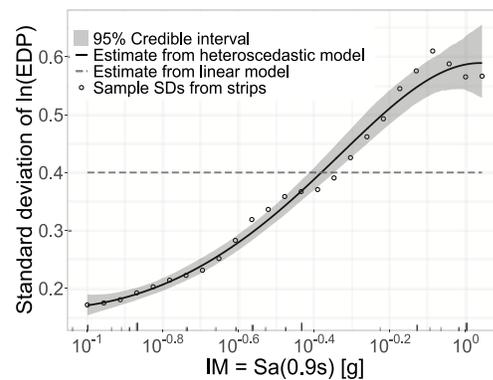
f) Standard deviation of $lnEDP|IM$ for pier 3

Figure 4 Comparison of mean and standard deviation for piers 1, 2, and 3 based on two approaches

Figure 4 shows a comparison of the mean and standard deviation obtained from the fitting of the two aforementioned models. From Figures 4 (a), (c), and (e), it can be seen that for the seismic demand means of the three piers, the fitting results of the heteroscedastic model fit the sample means relatively well, and the classical linear regression model results are slightly worse. By comparison, it was found that the sample means showed a slight nonlinearity overall. The non-linear variation trend to the demand means of piers 1 and 3 was identical owing to similar (symmetric) boundary conditions, while that for pier 2 was slightly different in comparison. Because the heteroscedastic model employed in this study selects the third-order polynomial as the basis function, it can fit the non-linear trend of the mean better with an increase in ground motion intensity. The classical linear regression model uses a first-order linear model for analysis resulting in straight line prediction, which fails to reflect the nonlinear tendency.

From Figures 4 (b), (d), and (f), it can be seen that the standard deviation increases with an increase in ground shaking intensity, but there are some differences in the specific change trends of the three piers. For piers 1 and 3, it is observed that the standard deviation sample has a peak value when the ground motion intensity is large, and after reaching the peak, there is a descending section (saturation section), rather than being strictly monotonous. This phenomenon is also consistent with the results obtained by Baker.[4] The heteroscedastic model fails to accurately fit this variation completely. However, the credible interval range obtained from the posterior distribution also increases with increasing ground motion intensity, reflecting the trend changes for uncertainty in this region. Therefore, the variance of estimates is also non-constant and exhibits some heteroscedasticity. For pier 2, the standard deviation sample fluctuates to a certain extent within the moderate zone of ground motion intensity, and then further increases. The heteroscedastic model describes the overall trend better. In contrast, the classical linear model has a constant standard deviation estimation due to the homoscedasticity assumption, which does not change anywhere. The estimates of the standard deviation for piers 1, 2, and 3 are 0.409, 0.340, and 0.401, respectively, and the results significantly deviate from the sample data. Therefore, the classical linear model cannot describe the fluctuation of the standard deviation within the studied ground motion intensity range, and thus leads to large errors. In general, the fitting results of the heteroscedastic model are closer to the sample data and better fit the variation pattern of the standard deviation.

In addition, the aforementioned results also show that for different components from the identical structure, even if the material and geometric properties are essentially the same, the differences in their boundary conditions can lead to inconsistencies in the statistical characteristics of the seismic demand and its variation pattern with ground motion intensity.

*4.2 Multivariate heteroscedasticity*

Based on the data obtained from the analysis in the previous section, a joint probabilistic seismic demand model for the three piers was developed in logarithmic space. As mentioned earlier, the covariance regression method proposed by Hoff et al. was used to establish the multivariate heteroscedasticity model, as shown in Equation 10. Similar to the aforementioned univariate heteroscedasticity model, herein, a third-order polynomial was used as the basis function for the regression, while the covariance was inscribed using a rank-3 model, as follows: Suppose $x_i = \ln IM_i$, $\boldsymbol{y_i} = (\ln EDP_i^1, \ln EDP_i^2, \ln EDP_i^3)^T$, and $\boldsymbol{x_i} = (1, x_i, x_i^2, x_i^3)^T$

Mean model: $\mathrm{E}(\boldsymbol{y_i}|x_i) = \boldsymbol{A}\boldsymbol{x_i}$

Covariance model: $\mathrm{Cov}(\boldsymbol{y_i}|x_i) = \boldsymbol{\Psi} + \boldsymbol{B_1}\boldsymbol{x_i}\boldsymbol{x_i}^T\boldsymbol{B_1}^T + \boldsymbol{B_2}\boldsymbol{x_i}\boldsymbol{x_i}^T\boldsymbol{B_2}^T + \boldsymbol{B_3}\boldsymbol{x_i}\boldsymbol{x_i}^T\boldsymbol{B_3}^T$

Here, $(\boldsymbol{A}, \boldsymbol{B_1}, \boldsymbol{B_2}, \boldsymbol{B_3}, \boldsymbol{\Psi})$ are the matrices of parameters to be estimated, in which $\boldsymbol{A}$ is the coefficient matrix of the mean with $p \times q$ dimension, $\boldsymbol{B_1}, \boldsymbol{B_2},$ and $\boldsymbol{B_3}$ are the coefficient matrices of the random effects with $p \times q$ dimension, and $\boldsymbol{\Psi}$ is the baseline covariance matrix with $p \times p$ dimension; herein, $p = 3$ is the number of outputs, and $q = 4$ is the number of covariates.

A semi-conjugate prior distribution can be set as the prior distribution for $(\boldsymbol{A}, \boldsymbol{B_1}, \boldsymbol{B_2}, \boldsymbol{B_3}, \boldsymbol{\Psi})$, in which $p(\boldsymbol{\Psi})$ is an inverse-Wishart distribution $\mathcal{W}^{-1}(\boldsymbol{\Psi_0}^{-1}, \nu_0)$ with expectation $\boldsymbol{\Psi_0}/(\nu_0 - p - 1)$, and $\boldsymbol{C} = (\boldsymbol{A}, \boldsymbol{B_1}, \boldsymbol{B_2}, \boldsymbol{B_3})$ is a matrix normal prior distribution, $\{\boldsymbol{C}|\boldsymbol{\Psi}\} \sim$ matrix normal$(\boldsymbol{C_0}, \boldsymbol{\Psi}, \boldsymbol{V_0})$. Subsequently, the Gibbs sampler proceeds by iteratively sampling the parameters from their full conditional distributions. In the absence of strong prior information, the default values for the prior parameters $\{\boldsymbol{C_0}, \boldsymbol{V_0}, \boldsymbol{\Psi_0}, \nu_0\}$ were used as recommended in the literature.[27] The 'covreg' package in R was used to implement the model fitting and obtain the Bayesian estimates through Gibbs sampling through the aforementioned priors.[35] An MCMC chain for 15000 iterations was run with thinning of every 10 samples, and the first 200 post-thinning samples were dropped as burn-in, and the trace plots of quantities for convergence were checked.

Meanwhile, the first-order multivariate linear model was selected as the comparison model. The mean and covariance models were $\mathrm{E}(\boldsymbol{y_i}|x_i) = \boldsymbol{\alpha_0} + \boldsymbol{\alpha_1}x_i$ and $\mathrm{Cov}(\boldsymbol{y_i}|x_i) = \boldsymbol{\Sigma}$, respectively. The parameters $(\boldsymbol{\alpha_0}, \boldsymbol{\alpha_1}, \boldsymbol{\Sigma})$ were estimated using the classical least squares method.

Because the regression results of the marginal distribution (mean and standard deviation) were similar to those of the univariate heteroscedastic model, these are repeated here. In this study, the analysis of the correlation of the logarithmic demands between different components was mainly focused. Therefore, the samples of correlation coefficients were extracted from the posterior distribution of the covariance matrix, and their variation patterns with the ground motion intensity measure are presented in Figure 5.

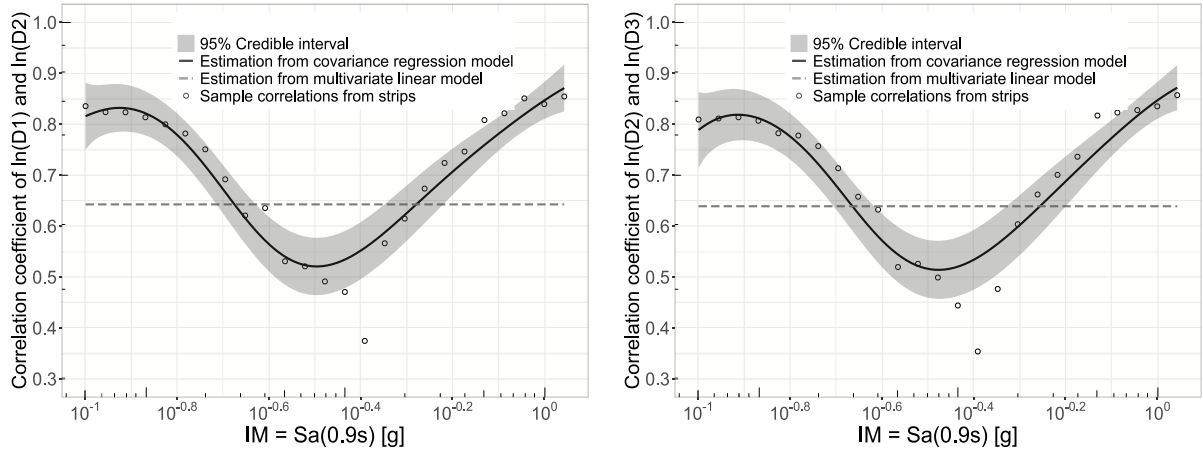

a) Correlation coefficient of $lnEDPs|IM$ for piers 1 and 2  b) Correlation coefficient of $lnEDPs|IM$ for piers 2 and 3

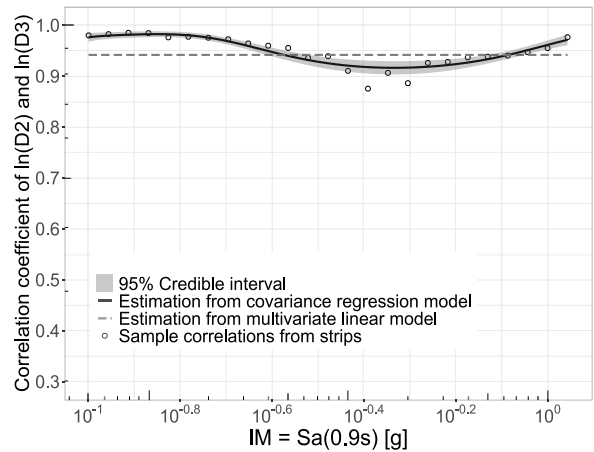

c) Correlation coefficient of $lnEDPs|IM$ for piers 1 and 3

Figure 5 Comparative illustration of correlation coefficients based on two approaches

Figure 5 shows a comparison of the correlation coefficients estimated by the two previous models. Here, $\rho_{12}$, $\rho_{23}$, and $\rho_{13}$ represent the correlation coefficients between the logarithmic seismic demands of the three piers in pairs, respectively. Contrasting to the standard deviation, the correlation coefficients did not change monotonically with ground motion intensity. Figure 5 shows that the variations in $\rho_{12}$ and $\rho_{23}$ are approximately the same. When the ground motion intensity is small, a relatively stable zone is observed. The correlation coefficient gradually decreases as the ground shaking intensity increases, reaching the minimum value when the ground shaking is $Sa \approx 0.4g$ ($lnSa = -0.9$); subsequently, the correlation coefficient increases as the ground shaking intensity increases. The scattered points are distributed in the interval of $(0.3, 0.9)$, and the general appearance is roughly "U-shaped." The trend of $\rho_{13}$ is approximately same as that of the first two. Because of the similarity (symmetry) in the boundary conditions of piers 1 and 3, the seismic response tends to be the same, resulting in the correlation coefficients of both always remaining at a high level within the studied seismic intensity range, and the correlation coefficient values are always greater than 0.85. The estimated results of the correlation coefficients obtained using the covariance regression model show a good fit overall; however, it cannot accurately represent the abrupt changes around $Sa = 0.4g$. The credible interval of $\rho_{13}$ fluctuates less with the variation in the ground motion intensity, while the credible intervals of $\rho_{12}$ and $\rho_{23}$ fluctuate relatively large. Because of the assumption of homoscedasticity, the estimated three correlation coefficients are 0.643, 0.639, and 0.942, which deviate more from the original sample data, and the estimated results cannot reflect the fluctuation of the correlation coefficients among the seismic responses of piers in the studied ground shaking intensity range. Overall, the fitting results using the covariance regression model were consistent with the actual situation.

The multivariate heteroscedasticity of the logarithmic seismic demands of the piers can also be presented in the form of prediction ellipses, as shown in Figure 6. In this study, for conciseness, only the results of five ground motion intensities are shown at equal intervals in pairs. It evident that the residuals of the logarithmic seismic demand of the piers are evenly distributed on both sides of the original point, regardless of the ground motion intensity. The more "flat" the shape of the prediction ellipse means the stronger correlation of the seismic response of the pier, and the more "rounded" the shape indicates a weaker correlation of the seismic responses of the piers. From Figure 6, it is observed that with an increase in the ground motion intensity, the shape of the prediction ellipse begins to change from "flat" to "round" and then slowly becomes "flat." This means that the correlation between piers and columns varies from high to low and again increases to high. From the residual prediction ellipses of piers 1 and 3, it can be observed that although these change with the ground motion intensity, their shape has always been "shuttle-shaped," which indicates that the two pier columns always maintain a high correlation, and the results are consistent with the variation pattern of the correlation coefficient estimation in Figure 5. The gray dashed ellipse lines are the 90% prediction ellipses calculated by the multivariate linear model, in which the center points are not located at the coordinate origin, indicating that there is a certain deviation in its estimation. Because of the assumption of homoscedasticity, the size and direction of the ellipses always remain unchanged, which is significantly different from the actual situation.

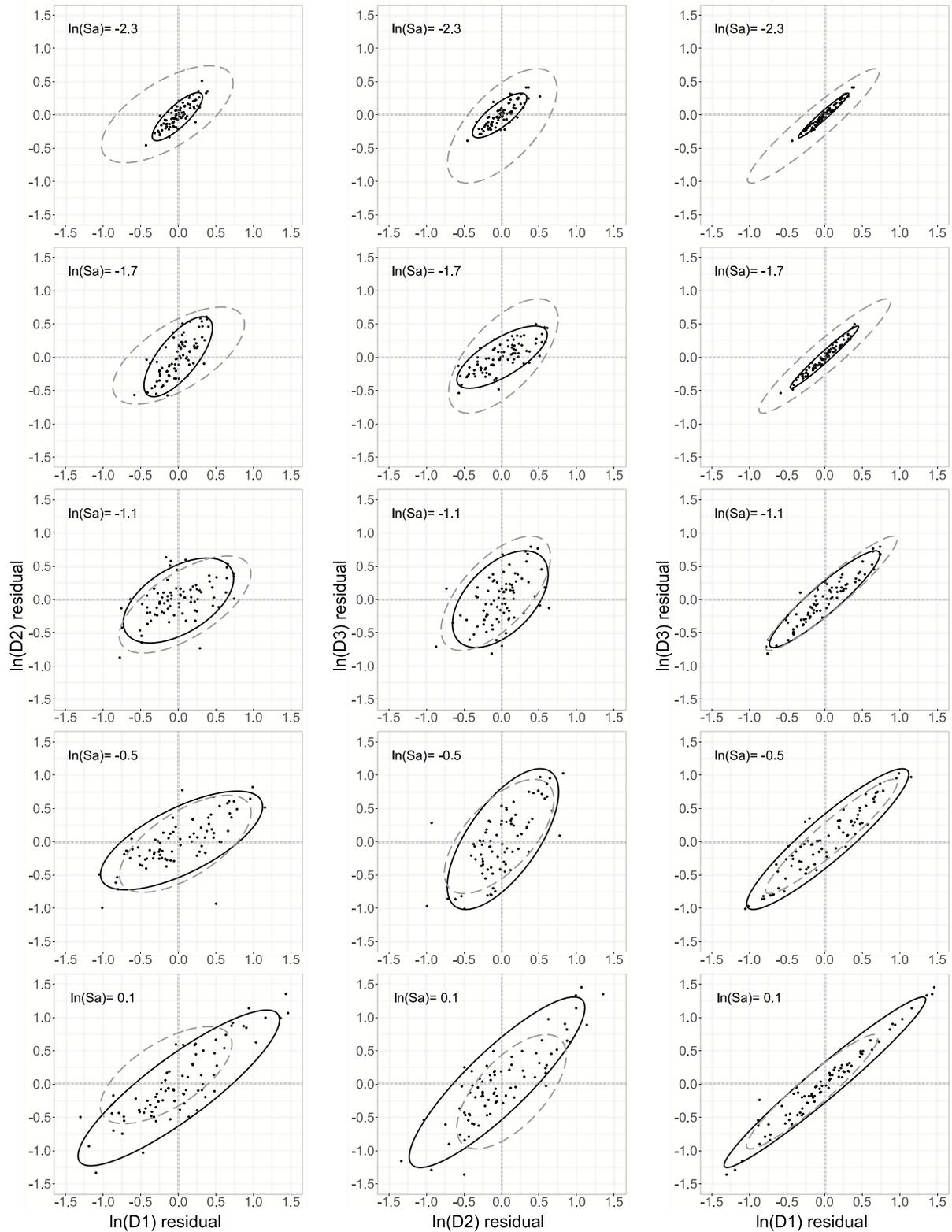

Figure 6 Residual data and 90% prediction ellipses for different IM. The dots represent the residuals of the logarithmic demand to the covariance regression model, and the solid and dotted ellipses correspond to the covariance regression and homoscedastic multivariate linear models, respectively

## 5. DISCUSSION AND CONCLUSION

Heteroscedasticity can arise due to numerous reasons, including, a strong nonlinear relationship between the dependent variable $Y$ and independent variable $X$, leading to inaccurate data transformations or incorrect functional form settings, or significant autocorrelation of $Y$ itself (mainly occurring in time series analysis, such as volatility clusters). In addition, there is a common possibility of variable omissions, that is, the residual includes some variables that are linearly related to $X$, but are not considered by the model, resulting in the correlation between the residuals and independent variables. An intuitive example is illustrated as follows: assume a real model $y_i = \beta_1 x_i + \beta_2 z_i + \varepsilon_i$, where $E(\varepsilon_i \mid x_i, z_i) = 0$, $Var(\varepsilon_i \mid x_i, z_i) = \sigma^2$, but $\rho(x_i, z_i) \neq 0$. For some reason, someone estimates a model such as $y_i = \gamma x_i + u_i$, where $u_i = \beta_2 z_i + \varepsilon_i$. Because a certain correlation exists between $x_i$ and $z_i$, thus leading to $E(u_i \mid x_i) \neq 0$ and $Var(u_i \mid x_i)$ is not a constant. It is worth clarifying that $E(u_i \mid x_i) \neq 0$ cannot be observed in reality, however, $Var(u_i \mid x_i)$ can be detected to be correlated with $x_i$ by the heteroscedasticity tests. Therefore, heteroscedasticity can be considered as a representation of epistemic uncertainty, because sometimes the real model cannot be determined.

The most likely explanation for the heteroscedasticity phenomenon in the probabilistic seismic demand analysis is that a scalar ground motion intensity measure cannot fully characterize all the features of the ground motion input, and some of the missing features may contribute to the maximum seismic demand of the structural component. To overcome this deficiency, a good idea is to add more explanatory variables (other ground motion intensity measures); however, it will face the problems of feature selection and collinearity of the predictor variables in practical operation. Furthermore, when the probabilistic seismic demand analysis is implanted into the PBEE framework, a connection problem between the front and rear models will be encountered. For instance, the joint probability distribution of multiple ground motion intensity measures cannot usually be provided by conventional probabilistic seismic hazard models. Therefore, to improve the accuracy of probabilistic seismic demand analysis in the conventionalized framework is to model and describe the uncertainty itself, as discussed earlier.

In this study, for inference methods, the Bayesian approach was selected as the statistical inference method for univariate and multivariate heteroscedasticity models. The major benefits are as following. First, it is a natural and logically self-consistent method for combining prior information with data, allowing past information about parameters to be combined, and forming prior distributions for future analysis. The previous posterior distribution can be used as a prior when new observations are available. Second, Bayesian analysis can directly estimate any function of the parameters without using the method of estimating the function by plugging the estimated parameters into generalized functions, while being able to provide very natural probabilistic interpretations of the results (e.g., the standard deviation has a 0.95 probability of falling within the 95% credible interval). Finally, it provides very flexible settings for a wide range of models, such as hierarchical and mixture models. Combined with the current widely used MCMC or other numerical methods, it is easier to handle the estimation and statistical inference of almost all parametric models. Its main disadvantage is that the produced posterior distributions could be strongly influenced by the prior distribution in some cases. Therefore, the selection of the prior distribution must be handled carefully; otherwise, it may produce misleading results. Some related principles have been summarized in literature.[36] In addition, it is usually accompanied by high computational costs, specifically in complex models with a large number of parameters. In this study, the classical linear regression was adopted for comparison using the least-squares method (the Frequentist method) due to its simplicity and ease of implementation, as well as the convergence with the conventional custom.

In terms of model setting, this study used the third-order polynomial as the basis function to fit the nonlinear parts of the mean and variance, and the rank of the random effect was considered as three in the covariance regression model. In the preliminary trial calculation, it was observed that adoption of a low-order scheme resulted in underfitting. The polynomial regression was equivalent to the Taylor expansion of the objective function in theory, and the more the considered terms, the higher the fitting accuracy for the objective. However, from past experiences, it was known that the order of polynomials should not be greater than 3 to prevent overfitting. The determination of the optimal order and rank of the random effect in both mean and variance or covariance models is still an emerging topic for researchers. Meanwhile, a more complicated relationship can be expressed using a non-linear function, such as a spline or Gaussian process (GP). The inference of continuous values with a Gaussian process prior is known as Gaussian process regression[37] or kriging, which is a Bayesian non-parametric method of interpolation based on a Gaussian process governed by prior covariance. In the standard Gaussian process regression model, the random residuals are assumed to be uncorrelated zero-mean Gaussian random variables with global or constant variance. For univariate heteroscedasticity, the noise variance can be modeled using a second GP in addition to the GP governing the noise-free output value,[38,39] which can be described as follows: $y_i|f_1(x), f_2(x), x_i \sim \mathcal{N}\left(f_1(x_i), \exp(f_2(x_i))\right)$, where $f_1(x) \sim \mathcal{GP}(0, k_1(\cdot,\cdot))$ and $f_2(x) \sim \mathcal{GP}(0, k_2(\cdot,\cdot))$. In addition, some researchers have proposed multivariate heteroscedasticity models based on Gaussian processes,[40,41] which can be used in the development of joint probabilistic seismic demand models in the future.

In terms of the analysis scheme, MSA was employed in this study for the sample generation of EDP-IM datasets. Its main purpose was to obtain the statistical characteristics from the samples of the strips under different ground motion intensities to facilitate the comparison with the estimation results. However, the proposed methods are also applicable to cloud analysis. In the specific implementation procedure, it should be ensured that the ground motion intensities of the selected ground motion records are evenly distributed within the range of ground motion intensity of interest, i.e., the number of samples under different ground motion intensities should be kept comparable to ensure the robustness of the regression results. In addition, the determination of the number of finite element samples required for constructing the heteroscedastic seismic demand model remains an open issue for both MSA or cloud analysis methods, i.e., obtaining plausible statistical results with minimal FEA cost deserves further investigation.

Owing to space limitations, the specific applications of the constructed models are not discussed quantitatively in this paper. As mentioned previously, further studies related to the effects of heteroscedastic models on the results of seismic fragility and risk (resilience) analysis may be performed subsequently. These issues can also be analyzed and discussed in Bayesian frameworks, with the major advantage that epistemic uncertainty can be considered. Seismic fragility, as one of the most direct applications of PSDM, has been widely studied in the last three decades. It is worth noting that, compared with the seismic fragility generated by the conventional first-order linear homoscedastic model, the errors due to the nonlinear mean might be larger than those due to heteroscedasticity in some cases. For a specific structural component, if the mean estimate is unbiased and the variance is monotonically increased, it can be concluded that the damage probability calculated through the heteroscedastic model would be smaller than that of the homoscedastic model at lower ground motion intensities, and the damage probability calculated by the heteroscedastic model would be larger than that of the homoscedastic model at higher ground motion intensities; while, at the intersection of the variance estimates, both are equivalent. Regarding the effect of heteroscedasticity on seismic risk (resilience), it essentially involves the propagation of uncertainty in the PBEE framework and is worth exploring using detailed simulation-based methods. For a structural system containing multiple components (e.g., a multi-span large bridge), the influence of the non-constant correlation of seismic demands may not be negligible for seismic risk or resilience assessment.

ACKNOWLEDGEMENT

The author thanks Prof. Paolo Gardoni of the Department of Civil and Environmental Engineering at the University of Illinois at Urbana-Champaign, whose classic paper published 20 years ago guided the author into the world of Bayesian statistics. This paper was conceived during the author's visiting scholarship at UIUC in 2018, and the author thanks him for his guidance, discussions, and help during the visit. The author also thanks his graduate students Wenfeng Lin and Huaitao Zhou, whose prior exploratory work provided the basis for the completion of this paper. The author thanks the National Natural Science Foundation of China and the Natural Science Foundation of Fujian Province (2020J01478) for their financial support.

Note: Only grayscale figures were used to minimize the cost of the printed paper.